\documentstyle[prl,aps,epsf]{revtex}         % 2 col mode

\tighten
\begin{document}
%
%\baselineskip=8.5mm  % preprint mode
%\draft               % preprint mode
% 2 col mode:
\twocolumn[\hsize\textwidth\columnwidth\hsize\csname@twocolumnfalse\endcsname
\title{Amplification of light and atoms in a Bose--Einstein condensate}

\author{S. Inouye, R. F. L\"ow, S. Gupta, T. Pfau,\\
A. G\"orlitz, T. L. Gustavson, 
D. E. Pritchard and W. Ketterle}

\address{Department of Physics and Research Laboratory of
Electronics, \\
Massachusetts Institute of Technology, Cambridge, MA 02139, USA}

\date{\today}
\maketitle

\begin{abstract}
A Bose-Einstein condensate illuminated by a single off-resonant laser beam 
(``dressed condensate'')
shows a high gain for matter waves and light. We have characterized the optical and atom-optical
properties of the dressed condensate by injecting light or atoms, 
illuminating the key role
of long-lived matter wave gratings produced by the condensate at rest and recoiling atoms. The
narrow bandwidth for optical gain gave rise to an extremely slow group 
velocity of an amplified light pulse ($\sim$~1~m/s).
\end{abstract}


\vskip1pc
] %2 col mode
% NEXT TWO LINES NOT REVTEX!

The field of atom optics is based on profound analogies between electromagnetic waves and matter
waves. Both light and atoms can be amplified by bosonic stimulation, and this has led to the
optical laser and the atom laser, respectively. Recently, Bose-Einstein 
condensates illuminated by
an off-resonant laser beam (``dressed condensates'') were used to realize phase-coherent
amplification of matter waves~\cite{MWA,Kozuma99}.  The amplification process involved the
scattering of a condensate atom and a laser photon into an atom in a recoil mode 
and a scattered photon.
This four-wave mixing process between
two electromagnetic fields and two Schr\"odinger fields
became a self-amplifying process above a threshold laser
intensity, leading to matter wave gain. 
However, the symmetry between light and atoms
indicates that a dressed condensate should not only amplify injected atoms, but also injected
light.

In this paper, we focus on the optical properties of a dressed condensate
above and below the threshold for the matter wave amplification.
The optical gain below the threshold has a narrow bandwidth
due to the long coherence time of a condensate.
This resulted in an extremely slow group velocity for
the amplified light. Above the threshold, we observed non-linear
behavior. The build-up of 
a macroscopic matter wave grating inside the condensate
is the direct consequence of phase-coherent amplification of matter waves and
was observed {\it in situ} by a pump-probe measurement.

Fig.\ 1 shows the basic light scattering process.  An atom scatters a photon from the laser beam
(called ``dressing beam'') into another mode and receives the corresponding
recoil momentum and energy.  Injection of atoms or light turns this 
{\it spontaneous} process into a
{\it stimulated} process. If atoms are injected, they form a matter wave grating (an
interference pattern with the condensate at rest) and this grating diffracts light. The diffraction
transfers recoil momentum and energy to the atoms, which results in a growth of the grating and therefore the
number of atoms in the recoil mode --- this is the intuitive picture for 
atom gain. If probe light is injected, it forms a moving standing wave with the dressing beam, and this
light grating diffracts atoms. This diffraction transfers photons into 
the probe beam mode, resulting in optical gain.

\begin{figure}[htbf]
\epsfxsize=69mm \centerline{\epsfbox{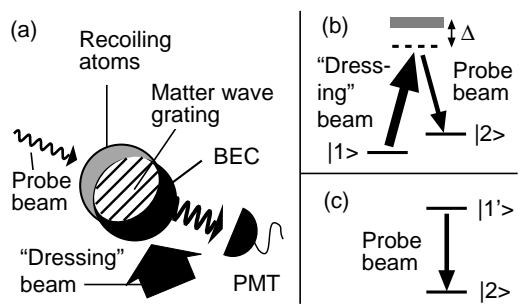}}\vspace{0.5cm}
\caption{Amplification of light and
atoms by off-resonant light scattering. (a) The fundamental process is the absorption of a photon from
the ``dressing'' beam by an atom in the condensate (state $|1 \rangle$), 
which is transferred to a
recoil state (state $|2 \rangle$) by emitting a photon into the probe field.  The intensity in the
probe light field was monitored by a photomultiplier. 
(b) The two-photon Raman-type transition between two 
motional states ($|1\rangle, |2\rangle$) gives rise to a narrow resonance.
(c) The dressed condensate is the upper state ($|1^{\prime}\rangle$) of 
a two-level system, and decays to the lower state (recoil state of 
atoms, $|2\rangle$) by emitting a photon.
}
\end{figure}

For low dressing beam intensity, the probe beam gain is due to 
the two-photon gain of individual atoms (Fig.\ 1b). 
By introducing the dressed atom picture~\cite{dress}, the optical gain can be 
understood as the gain of fully inverted two level system (Fig.\ 1c).
The atoms in the condensate and the photons in the dressing beam form a 
dressed condensate, corresponding to the upper state
 ($|1^{\prime}\rangle$). The lower state is the recoil state of atoms 
 ($|2\rangle$) and
the dressed condensate can ``decay'' to the lower state by emitting a 
photon into the probe beam mode.  A fully inverted two-level system
with dipole coupling would have a gain cross section of 
$6\pi\lambdabar^{2}$ for radiation with
wavelength $\lambda (= (2\pi)\lambdabar)$. For the Raman-type system in Fig.\ 1b, the gain is reduced by the excited
state fraction, $R_{\rm D} / \Gamma$ (where
$R_{\rm D}$ is the Rayleigh scattering rate for the dressing beam and $\Gamma$
is the linewidth of the single-photon atomic resonance)
and increased 
by $\Gamma/\Gamma_{2}$, the ratio of the
linewidths of the single-photon and two-photon Bragg resonances.
Thus the expected cross-section for gain is
\begin{equation}
\sigma_{\rm gain}=6\pi\lambdabar^{2} \frac{R_{\rm D}}{\Gamma_{2}} ,
\label{twophotongain}
\end{equation}
which is proportional to the intensity of the dressing beam.
A Bose--Einstein condensate has a very narrow
two-photon resonance width $\Gamma_{2}$ of only a few kHz.
The residual linewidth stems from the loss of overlap between the two
motional states and from the inhomogeneous density distribution~\cite{Bragg}.

A gain with narrow bandwidth causes a slow group velocity of light.
The gain represents the imaginary part of the complex index of refraction.
A sharp peak in the gain implies a steep dispersive shape for the real part 
of the index of refraction $n(\omega)$.
This results in a small value of the group velocity $v_{g}$
\begin{equation}
{v_{g}}=\frac{c}{n+\omega\frac{dn}{d\omega}}
\end{equation}
and in large pulse delays. More precisely, a narrow-band gain feature with 
gain $g$ and width
$\gamma$ leads to an amplified pulse with a delay time of 
$\tau_{\rm D}=(\ln{g})/ \gamma$.

\begin{figure}[htbf]
\epsfxsize=69mm \centerline{\epsfbox{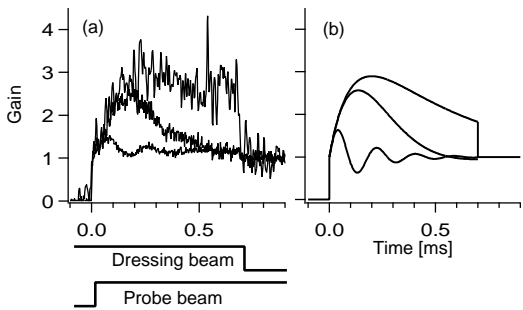}}\vspace{0.5cm} 
\caption{Gain and temporal behavior of light
pulses propagating through a dressed condensate. (a) Observed probe pulse output from a dressed condensate. The probe light
intensities were
$5.7 \,{\rm mW/cm}^{2}$ (bottom),
$1.5 \,{\rm mW/cm}^{2}$ (middle),
$0.10 \,{\rm mW/cm}^{2}$ (top), while the dressing beam
intensity was
$5 \, {\rm mW/cm}^{2}$, which was just below the threshold for superradiance.
The plotted signals were normalized by the
incident probe intensity and show the gain for the probe light.
(b) Calculated probe light output for typical
experimental parameters. Rabi oscillations develop into steady state gain 
as the intensity of the probe
light is reduced.
}
\end{figure}

For the experimental study of optical properties of a dressed condensate, elongated Bose--Einstein
condensates consisting of several million sodium atoms in the 
$F\!=\!1, m_{F}\!=\!-1$  state were prepared in a magnetic
trap~\cite{mom96}. The condensate was illuminated (``dressed'') with a single laser
beam that was red-detuned by 1.7~GHz from the $3S_{1/2}, F=1 \rightarrow 3P_{3/2}, F=0,1,2 $
transition. Both the dressing beam and the probe beam
were in the plane perpendicular to the long axis of the condensate,
and intersected at an angle of 135 degrees.
The probe beam was red-detuned from the dressing beam by $91 \, {\rm kHz}$, 
which was determined to be the
resonance frequency for the two-photon Bragg transition.
This small frequency difference between
the two light beams was realized by
deriving both beams from a common source, and then passing them 
through two separate acousto-optical modulators that were driven with
separate frequency synthesizers.
The probe beam, which propagated parallel to the axis of imaging, was a
few millimeters in diameter and much larger than the condensate,
which was $20 \, \mu$m in diameter and  $200 \, \mu$m in length.
In order to block all the light that did not pass through the 
condensate,  
a  $25\, \mu {\rm m} \times 100 \, \mu {\rm
m}$ slit was placed at an intermediate imaging plane where the condensate
was two times magnified. The
light transmitted by the slit was recorded with a photomultiplier. 
The polarization of each beam was set parallel to the long axis of the 
condensate to suppress superradiance to other recoil modes~\cite{super}.

The main results of this paper are shown in Figs.\ 2 and 3.
In order to measure the gain of the dressed condensate,
we used long square probe pulses
during which
the dressing beam was switched off (Fig.\ 2).
At the lowest probe intensity, the depletion of atoms in the condensate 
was negligible and a clear step at the switch off was observed,
corresponding to a gain of $\sim 2.8$.
The initial rise time $\sim 100 \, \mu s$
is the coherence time of the dressed condensate.

\begin{figure}[htbf]
\epsfxsize=69mm \centerline{\epsfbox{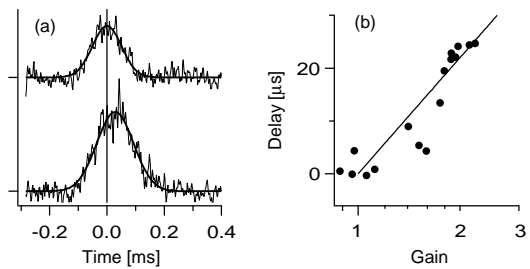}}
\vspace{0.5cm}
\caption{Pulse delay due to light amplification.
(a) About 20 $\mu$s delay was observed
when a Gaussian pulse of about 140 $\mu$s width and 0.11 mW/cm$^{2}$ peak intensity
was sent through the dressed condensate (bottom trace). The top trace
is a reference taken without the dressed condensate. Solid curves are 
Gaussian fits to guide the eyes.
(b) The observed delay $\tau_{\rm D}$ was proportional to $(\ln g)$, 
where $g$ is the observed gain.
}
\end{figure}

The square pulse
response observed in Fig.\ 2a already indicates long pulse delays.
The distortion of the pulse is due to the large frequency bandwidth contained in the
square pulse. This was avoided by modifying the pulse shape to be
Gaussian, but keeping the peak intensity of the probe beam at the same 
level. Fig.\ 3a shows that such pulses were
delayed by about 20 $\mu$s across the 20 $\mu$m wide condensate corresponding 
to a group velocity of 1 m/s. This is, to the best of 
our knowledge, one order of magnitude slower than any value published
thus far~\cite{Hau99}.

At high probe laser power we observed Rabi oscillations in the transmitted probe light (Fig.\ 2).
Note that all the traces were normalized by the probe beam intensity,
and the oscillatory trace at the bottom was obtained at the highest 
probe beam intensity.
They reflect simple two-level Rabi oscillations of atoms
between states $|1 \rangle$ and $|2 \rangle$ (Fig.\ 1b)
driven by the two-photon Bragg coupling.

The transition from Rabi oscillations to  steady state gain can be described by optical Bloch
equations. The two-level system $|1 \rangle$ and $|2 \rangle$ is coupled with the two-photon Rabi
frequency $\Omega =\Omega_{\rm D}\Omega_{\rm p}/2\Delta$
where $\Omega_{\rm D} (\Omega_{\rm p})$ is the Rabi frequency of the 
dressing (probe) beam and $\Delta$ is the detuning of the beams from 
the atomic resonance. The Bloch eqs.\ for atoms on the two-photon 
resonance take the following simple form:
\begin{eqnarray}
\dot{v}&=&-\frac{\Gamma_{2}}{2} v - \Omega w \label{vdot1}\\
\dot{w}&=&\Omega v \label{wdot}
\end{eqnarray}
where $v= 2 \;{\rm Im} (\rho_{12})$ represents the amplitude of
the matter wave grating($\rho_{ij}$ is the atomic density matrix)
and $w = \rho_{22} - \rho_{11}$ is the population difference between the two
states.

Fig.\ 2b shows the calculated probe output for a step function input.
Assuming constant $\Omega$ (valid for small optical gain), the solutions of the optical Bloch
equations show either Rabi oscillation ($\Omega \gg \Gamma_{2} /2$) or damped behavior ($\Omega \ll
\Gamma_{2} /2$). By reducing the probe power,
the Rabi oscillations slow down, and a (quasi-)steady state gain is obtained in the limit that they
are slower than the damping time. It is in this regime that the perturbative treatment with the
complex index of refraction applies.  Note that for longer times ($\gg \Gamma_{2}/ \Omega^2)$ the condensate
becomes depleted.

For large optical gain, the Rabi frequency $\Omega$ increases during the pulse and the above
treatment is no longer valid.  The population transfer to the recoil state ($\dot{w}$) results in
an increase of the number of the probe beam photons inside the condensate volume ($n_{\rm p}$)

\begin{equation}
\dot{n_{\rm p}}=\frac{c}{l} (n_{\rm p}^{0}-n_{\rm p}) + \frac{N_{0}}{2} 
\dot{w} \label{npdot}
\end{equation}
where $l$ is the length of the condensate with $N_{0}$ atoms and $cn_{\rm p}^{0}/l$ is the input photon flux. This
equation neglects propagation effects of the light by replacing the non-uniform electric field by
an average value~\cite{Haroche}. Using Eq.\ (\ref{wdot}) and replacing the photon number by the Rabi
frequency $\Omega^{2}=R_{D} 6\pi\lambdabar^{2} c n_{\rm p} / V$ ($V$ is the volume of the
condensate), we obtain
\begin{equation}
\dot{\Omega}=\frac{c}{l} ( \Omega^{0} - \Omega + \frac{G}{2} v ) 
\label{omegadot}
\end{equation}
where $\Omega_{0}$ is the two-photon Rabi frequency due to the input probe beam and the dressing
beam, and $G$ is defined by $G=N_{0}\; R_{\rm D}\; 6 \pi \lambdabar^{2} / 2A$ ($A$ is the
cross-section of the condensate perpendicular to the probe beam). The coupled 
equations (\ref{vdot1}) and (\ref{omegadot})
between the light and matter wave grating are analogous to the optical laser, where the atomic
polarization and the electric field inside the cavity are coupled. However, the role of atoms and
light is  reversed:  in the optical laser, the cavity lifetime is usually longer than the coherence
time of the atomic polarization, whereas in our case the extremely long coherence time of the
condensate dominates.

Adiabatic elimination of the light ($\dot{\Omega}=0$ in Eq.\ 
(\ref{omegadot})) and neglecting condensate
depletion ($w = -1$) lead to
\begin{equation}
\dot{v} = \frac{G-\Gamma_{2}}{2} v + \Omega^{0} \label{vdot2}
\end{equation}

Above the threshold for superradiance, ($G\ge \Gamma_{2}$), the matter wave grating builds up
exponentially. Below the threshold, it relaxes to $v=2\Omega_{0}/(\Gamma_{2}-G)$ with a time
constant of $2/(\Gamma_{2}-G)$, providing a gain for the probe beam of
$\Gamma_{2}/(\Gamma_{2}-G)$. This gain can be written as
\begin{equation}
1+\frac{G} {\Gamma_2-G}=1+ \frac{n_{0}\sigma_{\rm gain} l}{2}
\frac{\Gamma_2}{\Gamma_2-G} \label{gain}
\end{equation}
where $n_0$ is the condensate density.  In the low intensity limit, Eq.\ 
(\ref{gain}) reproduces the
two-photon gain discussed above (Eq.\ (\ref{twophotongain})).  Eq.\ (\ref{gain}) shows that the  effect of the coupled
equations is to replace the two-photon linewidth $\Gamma_2$ in Eq.\ 
(\ref{wdot}) by the ``dynamic''
coherence decay rate $\Gamma_2-G$.
The expansion of the optical gain $\Gamma_2/(\Gamma_2-G)= 1+ (G/\Gamma_2) + (G/\Gamma_2)^2 +
\ldots $ shows the transition from (linear) single-atom gain to (non-linear) collective gain.  At
the onset of superradiance, the optical gain diverges.

We studied this transition by first creating a matter wave grating with a Bragg pulse (using the
dressing and probe beams), and then observing its time evolution by monitoring the diffracted
dressing beam. The initial seed pulse was 100 $\mu$s long and transferred about 5\% of the atoms to
the recoil state.

At lower intensities for which collective effects were negligible, the grating showed a simple decay
(Fig.\ 4a)~\cite{decay}. At higher intensities, collective gain started to compensate the loss, and
at intensities above a threshold, net amplification was observed.  The 
initial growth rate (Fig.\ 4b)
followed the linear dependence on the intensity of the dressing beam ($\propto (G-\Gamma_2$))
predicted by Eq.\ (\ref{vdot2}) and Refs.~\cite{super,Meys99}.  The net growth of the matter wave grating corresponds
to atom amplification which was studied previously by observing an increase in the number of
recoiling atoms in time-of-flight images~\cite{MWA}.  Here we have monitored the dynamics of
amplification {\it in situ} by observing light instead of atoms.

Extrapolating the decay rate in Fig.\ 4b to zero intensity of the dressing 
beam, we obtained the
decay time of the matter wave grating $\Gamma_{2}$ of $100 \, \mu$s, in 
fair agreement with
the linewidth of the Bragg excitation process observed 
previously~\cite{Bragg}.

\begin{figure}[htbf]
\epsfxsize=79mm \centerline{\epsfbox{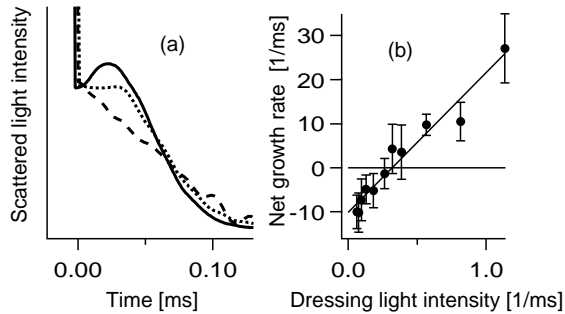}}\vspace{0.5cm}
\caption{Pump-probe spectroscopy of
a matter wave grating inside the condensate. (a) First, approximately 
5~\% of atoms were
transferred to the recoil state by the two-photon Bragg transition. Then the dynamics of the matter
wave grating was observed {\it in situ} by illuminating the grating with off-resonant light and
monitoring the diffracted light intensity. All traces were normalized to the same diffracted light
intensity at $t=0$. The dressing beam intensity was $2.9 \, {\rm mW/cm^{2}}$(bottom),
$5.7 \, {\rm mW/cm^{2}}$(middle), $13 \, {\rm mW/cm^{2}}$(top). (b) The initial growth rate of the
grating vs.\ light intensity shows the threshold for net gain. The intensity of the dressing beam
is given in units of the single-atom Rayleigh scattering rate. }
\end{figure}

Recent demonstrations of slow group velocities for light focused on electromagnetically induced transparency
in a three-level lambda system~\cite{Hau99}. This system features a narrow dip in a broad
absorption feature. In our system, the broad absorption line is 
missing.  Since the propagation of
resonant laser pulses is mainly determined by the narrow feature, both systems show analogous
behavior. Note that the three-level system shown in Fig.\ 1b does not have a dark state because it
has off-resonant couplings to other momentum states (not shown in the figure).
In the past, ``superluminal'' pulse propagation 
was observed in non-inverted, absorptive two-level systems~\cite{Chu82}.
Our scheme realizes the opposite case of gain and slow pulse 
propagation~\cite{Yariv71}.

The dressed condensate studied in this paper is a clean, model system 
for discussing optical and
atom-optical properties.
The optical amplification can be described as a reflection of the dressing light by a
matter wave grating. The initial delay time in the amplification of optical pulses is the time
necessary to build up the (quasi-) steady-state matter wave grating. The trailing edge of the
transmitted light pulse reflects the slow decay of quasi-particles.  Thus, the slow speed of light
is simply related to the build-up and decay of quasi-particles which we were able to monitor
directly.

The optical gain studied above clearly showed the transition from  single atom gain to collective
gain.  Previously, recoil related gain based on single atom phenomena (Recoil Induced Resonances),
was observed in cold cesium atoms~\cite{RIR}. Collective gain due to the formation of a density
grating was discussed as a possible gain mechanism for lasing action~\cite{CARL} (named
CARL---Coherent Atomic Recoil Laser) and pursued experimentally~\cite{CARLexp} with ambiguous
results (see~\cite{CARLquestion} and the discussion in~\cite{New_CARL}). Our
experiments clearly identify the two regimes and their relationship.

Our observation of the decay of the matter-wave grating can be regarded as pump-probe spectroscopy
of quasi-particles in the condensate.  The seeding Bragg pulse created the 
quasi-particles
(in this case condensate excitations in the free-particle regime).
One can control the momentum of the excited quasi-particles by the angle
between the laser beams.  This can also be used to excite phonon-like quasiparticles 
~\cite{phonon}. The pump-probe scheme presented here could be directly 
applied to the study of their lifetimes.

In conclusion, we have characterized the optical and atom-optical properties of a dressed
condensate.  The simple process of Rayleigh scattering gave rise to a rich variety of phenomena
including steady-state gain, Rabi oscillations, collective gain, and slow group velocities. We have
also introduced a pump-probe technique to study the lifetime of quasi-particles in a condensate.

We thank D. M. Stamper-Kurn for critical reading of the manuscript
and A. P. Chikkatur for experimental assistance.
This work was supported by the ONR, NSF,
ARO, NASA, and the David and Lucile Packard Foundation. 
A.G. would like to acknowledge support from DAAD and T.P. from the Alexander von Humboldt--Foundation.

\bibliographystyle{prsty}

\begin{thebibliography}{99}
\bibitem{decay}
   Because the lineshape of the Bragg resonance for the condensate atoms
   is not Lorentzian, the decay curve is not strictly exponential.
\end{thebibliography}

\end{document}